\documentclass[12pt]{article}

\begin{document}

\renewcommand{\theequation}
{\thesection.\arabic{equation}}
\thispagestyle{empty}
\vspace*{20mm} 
\begin{center}
{\LARGE {\bf Conformal Anomalies in Noncommutative}} \\
 \qquad \\
{\LARGE {\bf Gauge Theories}} \\

\vspace*{20mm}

\renewcommand{\thefootnote}{\fnsymbol{footnote}}

{\Large Tadahito NAKAJIMA 
\footnote[1]{E-mail address: nakajima@phys.ge.cst.nihon-u.ac.jp}} \\
\vspace*{20mm}

{\large {\it Laboratory of Physics, College of Science and Technology, 
Nihon University}} \\
{\large {\it Narashino-dai, Funabashi, Chiba 274-8501, Japan}} \\ 

\vspace*{20mm}

{\bf Abstract} \\

\end{center}

We calculate conformal anomalies in noncommutative gauge theories by using the 
path integral method (Fujikawa's method). Along with the axial anomalies and 
chiral gauge anomalies, conformal anomalies take the form of the 
straightforward Moyal deformation in the corresponding conformal anomalies in 
ordinary gauge theories. However, the Moyal star product leads to the 
difference in the coefficient of the conformal anomalies between 
noncommutative gauge theories and ordinary gauge theories. 
The $\beta$ (Callan-Symanzik) functions which are evaluated from the 
coefficient of the conformal anomalies coincide with the result of 
perturbative analysis.

\vspace*{15mm}

\clearpage
\section{Introduction}
\setcounter{page}{1}
\setcounter{equation}{0}

Gauge theories on noncommutative space-time (noncommutative gauge theories) 
have recently attracted much attention (for a review, see \cite{MRDNAN}). 
This is partly due to the realization that such theories actually occur in 
string theory with a constant NS-NS two-form field \cite{NSEW}. 
The noncommutative character of the Moyal star product leads to 
noncommutative gauge theories. For example, noncommutative $U(1)$ gauge 
theories have a character similar to ordinary non-Abelian gauge theories, 
although the gauge group is commutative. It is shown from perturbative 
analysis of the $\beta$ function that the noncommutative $U(1)$ Yang--Mills 
theory is asymptotically free \cite{CPMDSR, SMMVR}. 
An intrinsic feature of noncommutative gauge theories is 
the so-called UV/IR mixing \cite{SMMVR}. The planar diagrams controlled the UV 
properties, while nonplanar diagrams generally lead to new IR phenomena 
through the mixing.

Axial anomalies and chiral gauge anomalies have been actively studied in 
noncommutative gauge theories \cite{FANS0}-\cite{KIJK}. These 
anomalies can be calculated by perturbative analysis and the path integral 
formulation (Fujikawa's method). Chiral gauge anomalies can also be described 
using generalized descent equations \cite{LBMSAT}. It is known that these 
anomalies take the form of a straightforward Moyal deformation in the 
corresponding anomalies in ordinary gauge theories. However, this 
modification includes physical consequences 
to the chiral gauge anomalies. The noncommutative character of the Moyal star 
product actually leads to more restrictive conditions for anomaly 
cancellation \cite{JMGBCPM, LBMSAT}. The noncommutative chiral gauge 
theories with fermions in the fundamental representation 
are anomalous. The chiral gauge anomalies only come from planar diagrams in 
this representation. 
On the other hand, the noncommutative chiral gauge theories with fermions in 
the adjoint representation are anomaly-free (in four dimensions) 
\cite{CPM, FANS, JNMA}.  
Although not only planar diagrams but also nonplanar diagrams contribute to 
the chiral gauge anomalies in this representation, they cancel in each sector. 
Therefore, nonplanar diagrams do not contribute to the chiral gauge 
anomalies (in a single gauge group) \cite{CPM, KIJK}.

The noncommutative field theories include the noncommutativity parameter 
$\theta$ of dimension $[{\rm length}]^{2}$. Therefore, it is expected that 
the scale (or dilatation) invariance of the field theories is broken at the 
classical level even if the field theories are massless field theories. The 
breaking for scale invariance at the classical level was actually 
investigated in the Moyal deformed massless scalar field theory. 
In the classical scalar field theory, 
the variation of the action under the infinitesimal scale 
transformation is proportional to the change in the noncommutativity 
parameter induced by infinitesimal scale transformation \cite{AGJGHGLPMSRW}. 
Therefore, the Moyal deformed massless scalar field theory is invariant 
under the scale transformation including the change in the noncommutativity 
parameter. 
On the other hand, the Weyl symmetry, which is closely related to the scale 
invariance, is broken as a result of quantum corrections in the ordinary field 
theories. This phenomenon is well known as 
conformal (or Weyl) anomalies. It is an interesting problem to study how 
conformal anomalies are deformed by the Moyal star products in the 
noncommutative field theories.

In this paper, we have calculated conformal 
anomalies in four dimensional noncommutative gauge theories (on flat space) 
with fermions in a fundamental representation. Variants of 
the path integral method (Fujikawa's method) will be found to be 
suited for the calculations. The calculation in the path integral method is 
simple, although we needs some knowledge of Weyl 
transformations and breaking. 
We advance calculation of the conformal anomaly to Abelian gauge theory, 
the noncommutative QED, first. The generalization to 
non-Abelian gauge theory, the noncommutative 
QCD, is straightforward. The paper is organized as follows. 
In Sec. 2, we state the method of calculation for the conformal 
anomaly in the path integral method based on Ref. \cite{KF} after 
introducing the background field method for noncommutative QED. In Sec. 3, we 
calculate the conformal anomaly in noncommutative QED at the one-loop level. 
In ordinary gauge theories, 
there is a relation between the conformal anomaly and the $\beta$ function. 
Based on the relation, we evaluate the $\beta$ function in noncommutative 
QED and compare with the result from perturbative analysis. 
The calculating method shown in Sec. 3 is generalizably 
straightforward in the noncommutative QCD. It is stated by Sec. 4. 
Sec. 5 is devoted to a summary and discussion. 

%
%
\section{The background field method for noncommutative QED} 
\setcounter{equation}{0}
\addtocounter{enumi}{1}

Noncommutative gauge theories can be obtained by replacing the ordinary 
products of fields in the actions of their commutative counterparts by the 
Moyal star products, 
\begin{eqnarray}
f(x) \ast g(x) \!\!&=&\!\! \left.
e^{\frac{i}{2}\theta^{\mu\nu}
\frac{\partial}{\partial \xi_{\mu}}\frac{\partial}{\partial \zeta_{\nu}}}
f(x+\xi)g(x+\zeta) \right|_{\xi=\zeta=0} \nonumber \\
\!\!&=&\!\! \int \frac{d^{4}p}{(2\pi)^{4}}\int \frac{d^{4}q}{(2\pi)^{4}}
e^{-\frac{i}{2}p_{\mu}\theta^{\mu\nu}q_{\nu}}
e^{i(p_{\mu}+q_{\mu})x^{\mu}} \hat{f}(p)\hat{g}(q) \;,
\end{eqnarray}
%
where $\theta^{\mu\nu}=-\theta^{\nu\mu}$ is an antisymmetric real matrix. 
It is known that the matrix $\theta^{\mu\nu}$ is constrained by imposing 
unitarity on a noncommutative quantum field theories. The only allowed types 
of the matrix $\theta^{\mu\nu}$ are spacelike and lightlike 
\cite{JGMKTMMRSS}.

We begin with the noncommutative $U(1)$ Yang--Mills action, 
\begin{eqnarray}
S_{gauge}[A_{\mu}] = -\frac{1}{4g^{2}} \int d^{4}x \, 
F_{\mu\nu}(x) \ast F^{\mu\nu}(x) \;.
\end{eqnarray}
%
Here the field strength $F_{\mu\nu}(x)$ is 
\begin{eqnarray}
F_{\mu\nu}(x) = \partial_{\mu}A_{\nu}(x) - \partial_{\nu}A_{\mu}(x) 
-i[\,A_{\mu}(x), A_{\nu}(x) \,]_{M} \;, 
\end{eqnarray}
%
with the Moyal bracket $[\,A, B \,]_{M} = A \ast B - B \ast A$. 
The action (2.2) is invariant under the infinitesimal gauge transformation 
$\delta A_{\mu}(x)= \partial_{\mu} \lambda(x) 
-i[\, A_{\mu}(x), \lambda(x) \,]_{M}$ with the infinitesimal gauge 
transformation parameter $\lambda(x)$.

In order to compute the effective action including the quantum effect 
of the gauge field, we introduce the background field method for 
noncommutative gauge theories \cite{VVKGT, CPMFRR}. We decompose 
the gauge field $A_{\mu}$ into a background field $B_{\mu}$ and a fluctuating 
field $a_{\mu}$,
\begin{eqnarray}
A_{\mu}(x) = B_{\mu}(x) + a_{\mu}(x) \;.
\end{eqnarray}
%
Then the field strength decomposes as follows: 
\begin{eqnarray}
F_{\mu\nu} = F_{\mu\nu}[B] + D_{\mu}[B]a_{\nu} - D_{\nu}[B]a_{\mu} 
-i[\,a_{\mu}, a_{\nu}\,]_{M} \;,
\end{eqnarray}
%
where $F_{\mu\nu}[B]$ is the field strength of the background field $B_{\mu}$,  and $D_{\mu}[B]$ is the covariant derivative acting on the fluctuating 
field $a_{\mu}$, 
\begin{eqnarray}
F_{\mu\nu}[B](x) \!\!&\equiv&\!\! \partial_{\mu}B_{\nu}(x) - \partial_{\nu}B_{\mu}(x) 
-i[\,B_{\mu}(x), B_{\nu}(x) \,]_{M} \;, \\
 & & \nonumber \\
D_{\mu}[B]a_{\nu}(x) \!\!&\equiv&\!\! \partial_{\mu}a_{\nu}(x) 
-i[\,B_{\mu}(x), a_{\nu}(x) \,]_{M} \;.
\end{eqnarray}
%
Substitution in the action (2.2) of the field strength (2.5) and 
integration by parts yields
\begin{eqnarray}
S_{gauge}[a_{\mu}, B_{\mu}] \!\!&=&\!\! -\frac{1}{4g^{2}} \int d^{4}x \,
\biggl\{\; \biggr. F_{\mu\nu}[B] \ast F^{\mu\nu}[B] 
+ 4i a^{\mu} \ast [\; F_{\mu\nu}[B], \; a^{\nu} \;]_{M}   
\nonumber \\
& & -2a_{\mu} \ast D_{\nu}[B]D^{\nu}[B]a^{\mu} 
-2D_{\mu}[B]a^{\mu} \ast D_{\nu}[B]a^{\nu} + {\cal O}(a_{\mu}^{3}) 
\biggl. \; \biggr\} \;.
\end{eqnarray}
%
In deriving this action, we have used the classical 
equation of motion for the background field $B_{\mu}$. If the background 
field $B_{\mu}$ is regarded as fixed, the action (2.8) has the following 
local symmetry: 
\begin{eqnarray}
\delta a_{\mu}(x) = D_{\mu}[B] \lambda(x) 
-i [\;a_{\mu}(x), \; \lambda(x) \;]_{M} \;.  
\end{eqnarray}
%
In order to define 
the functional integral, we need to perform the gauge fixing for the local 
gauge symmetry implemented by the transformation (2.9). We choose a gauge 
fixing (GF) term and Faddeev--Popov (FP) ghost term 
(in the 't Hooft--Feynman gauge) as follows: 
\begin{eqnarray}
S_{GF + FP}[a_{\mu}, \; c, \; \bar{c}] \!\!&=&\!\! -\frac{1}{2g^{2}} 
\int d^{4}x \, \biggl\{\; \biggr. D_{\mu}[B]a^{\mu} \ast D_{\nu}[B]a^{\nu} 
\nonumber \\
& & -i \bar{c} \ast D^{\mu}[B] 
(\; D_{\mu}[B]c-i[\;a_{\mu}, c \;]_{M} \;)
\biggl. \; \biggr\} \;,
\end{eqnarray}
%
where $c(x)$ and $\bar{c}(x)$ are the ghost fields. Here the gauge fixing 
condition has been taken to be covariant with respect to the background field. 
We can obtain the gauge fixed action for the fluctuating field $a_{\mu}$ by 
adding Eqs. (2.8) and (2.10).

We next introduce the matter fields. The action for the massless fermion 
interacting with a background $U(1)$ gauge field $B_{\mu}$ is given by 
\begin{eqnarray}
S_{matter}[\bar{\psi}, \psi] = \int d^{4}x \, \bar{\psi}(x) \ast 
(i /\!\!\!\!D[B]) \psi (x) \;, 
\end{eqnarray}
%
where the covariant derivative acting on the fermion is defined 
by \cite{JMGBCPM, CPM, FANS} 
\begin{eqnarray}
/\!\!\!\!D[B] \psi(x) 
\equiv \gamma^{\mu} \partial_{\mu}\psi(x) -i\gamma^{\mu}B_{\mu}(x) \ast 
\psi(x)\;. 
\end{eqnarray}
%
The gauge-fixed action in the noncommutative QED is given by 
$S_{gauge} + S_{GF + FP} + S_{matter}$. Note that the gauge fixed action is 
still invariant under a local transformation:
\begin{eqnarray}
& & \delta \psi(x) = -ig \lambda(x) \ast \psi(x)\;, \qquad 
\delta \bar{\psi}(x) = +ig \bar{\psi}(x) \ast \lambda(x)\;, 
\nonumber \\
& & \nonumber \\
& & \delta B_{\mu}(x) = \partial_{\mu}\lambda(x) - 
 i [\;B_{\mu}(x), \; \lambda(x) \;]_{M}
\;, \qquad 
\delta a_{\mu}(x) = i [\;a_{\mu}(x), \; \lambda(x) \;]_{M} \;, \\
& & \nonumber \\
& & \delta c(x) = i [\;c(x), \; \lambda(x) \;]_{M} \;, \qquad 
\delta \bar{c}(x) = i [\;\bar{c}(x), \; \lambda(x) \;]_{M} \;. \nonumber 
\end{eqnarray}
%
The Wilsonian effective action is obtained by functional integration over the 
fluctuating field. The one-loop effective action $W[B]$ for the background 
field $B_{\mu}$ can be written as
\begin{eqnarray}
\exp(-W[B]) = 
\int {\cal D}\psi {\cal D}\bar{\psi}{\cal D}a_{\mu} {\cal D}c {\cal D}\bar{c}
\,\exp \left( \,
S_{quad}[a_{\mu}] + S_{quad}[c, \; \bar{c}] + S_{matter}[\bar{\psi}, \psi] 
\, \right)\;,
\end{eqnarray}
%
with 
\begin{eqnarray}
S_{quad}[a_{\mu}] \!\!&=&\!\! \frac{1}{2g^{2}} \int d^{4}x \,
\left\{\; a_{\mu} \ast D_{\nu}[B]D^{\nu}[B] a^{\mu}    
-2i a_{\mu} \ast [\; F^{\mu\nu}[B], \; a_{\nu} \;]_{M} \; \right\} \;, \\
& & \nonumber \\
S_{quad}[c, \; \bar{c}]  \!\!&=&\!\! \frac{1}{4g^{2}} \int d^{4}x \,
\biggl\{\; \biggr. i\bar{c} \ast D_{\mu}[B]D^{\mu}[B] c 
\biggl. \; \biggr\} \;.
\end{eqnarray}
%
Here, we perform a Wick rotation into Euclidean space-time with the metric 
$g_{\mu\nu}=-\delta_{\mu\nu}$ for the actual calculations.

In ordinary gauge theories, the one-loop conformal (or Weyl) anomalies can be 
simply evaluated by using the background field method in the path integral 
approach \cite{KF}. In this approach, the conformal anomalies are 
characterized as the Jacobian for the functional measure 
${\cal D}\hat{\psi} {\cal D}\hat{\bar{\psi}}
{\cal D}\hat{a}_{\mu} {\cal D}\hat{c} {\cal D}\hat{\bar{c}}$ 
with the field variables in flat space, 
\begin{eqnarray}
\begin{array}{ll}
\hat{\psi}(x) \equiv \sqrt[\scriptstyle 4]{|g|}\psi(x) (= \psi(x)) \;, 
& \hat{\bar{\psi}}(x) \equiv \sqrt[\scriptstyle 4]{|g|}\bar{\psi}(x) 
(= \bar{\psi}(x)) \;, \\ 
& \\
\hat{a}_{\mu}(x) \equiv \sqrt[\scriptstyle 4]{|g|}e^{i}_{\mu}a_{i}(x) 
(= a_{i}(x)) \;, 
& \\
& \\
\hat{c}(x) \equiv \sqrt[\scriptstyle 4]{|g|}c(x) (= c(x)) \;, 
& \hat{\bar{c}}(x) \equiv \sqrt[\scriptstyle 4]{|g|}\bar{c}(x) 
(= \bar{c}(x)) \;, 
\end{array}
\end{eqnarray}
%
respectively. Here $g$ is the determinant of the metric and $e^{i}_{\mu}$ is 
the vielbein in flat (Euclidean) space. 
The scale transformation can be regarded as a combination of the Weyl 
transformation and the coordinates transformation. 
The choice of the functional measure 
is dictated by the manifest covariance under the (general) coordinate 
transformation in curved space. Note that the redefinition of the field 
variables modifies Weyl transformations laws. We take the following 
transformation laws as Weyl transformation laws on noncommutative space: 
\begin{eqnarray}
\begin{array}{ll}
\psi(x) \rightarrow  \widetilde{\psi}(x) 
= \exp\left( -\frac{1}{2}\alpha(x) \right) \ast \psi(x) \;, 
& \bar{\psi}(x) \rightarrow  \widetilde{\bar{\psi}}(x) 
= \bar{\psi}(x) \ast \exp\left( -\frac{1}{2}\alpha(x) \right) \;, \\
& \\
a_{\mu}(x) \rightarrow \widetilde{a}_{\mu}(x) 
= \exp\left( -\alpha(x) \right) \ast a_{\mu}(x) \;, 
& \\
& \\
c(x) \rightarrow \widetilde{c}(x) 
= c(x) \ast \exp\left( -2\alpha(x) \right) \;, 
& \bar{c}(x) \rightarrow  \widetilde{\bar{c}}(x) 
= \bar{c}(x) \;, 
\end{array}
\end{eqnarray}
%
where $\alpha(x)$ is an infinitesimal arbitrary function. When the Moyal star  
products in Eqs. (2.18) are restored to the ordinary (commutative) products, 
the transformation laws (2.18) are also restored to the ordinary Weyl 
transformation laws for the field variables (2.17). In the next section, 
we derive the conformal anomaly in noncommutative QED 
(in the flat space limit) on the basis of the path integral approach. 
For this purpose, we will 
evaluate the associated Jacobian of functional measure in Eq. (2.14) under 
the transformation laws (2.18). For convenience, 
however, we suppose $\alpha(x)$ is an infinitesimal arbitrary constant 
hereafter. Namely, we treat the global Weyl transformations. 

%
%
%
\section{The conformal anomaly in noncommutative QED}
\subsection{The contribution from matter fields}
\setcounter{equation}{0}
\addtocounter{enumi}{1}

We first evaluate the contribution from the matter fields to the conformal 
anomaly. The global Weyl transformation laws for the matter fields are 
given by 
\begin{eqnarray}
\psi(x) & \longrightarrow & \widetilde{\psi}(x) 
= \exp\left( -\frac{1}{2}\alpha \right) \psi(x) \;, \nonumber \\ 
 & &  \\
\bar{\psi}(x) & \longrightarrow & \widetilde{\bar{\psi}}(x) 
= \bar{\psi}(x) \exp\left( -\frac{1}{2}\alpha \right) \;, \nonumber 
\end{eqnarray}
%
where $\alpha$ is a constant parameter. In order to define the integral 
measure of the fermionic fields more accurately, we decompose $\psi(x)$ 
and $\bar{\psi}(x)$ into eigenfunctions of the Dirac operator defined in 
Eq. (2.12),
\begin{eqnarray}
\psi(x) = \sum_{n}a_{n}\varphi_{n}(x) \;, \qquad 
\bar{\psi}(x) = \sum_{n}\bar{b}_{n}\varphi_{n}^{\dagger}(x) \;.
\end{eqnarray}
%
The coefficients $a_{n}$ and $\bar{b}_{n}$ are Grassmann numbers. 
The Dirac operator $/\!\!\!\!D[B]$ has real eigenvalues $\lambda_{n}$
\begin{eqnarray}
/\!\!\!\!D[B] \varphi_{n}(x) = \lambda_{n}\varphi_{n}(x) \;,
\end{eqnarray}
%
and the set of eigenfunctions $\{\varphi_{n}(x)\}$ is complete. 
We assume that the set of eigenfunctions is orthonormal: 
\begin{eqnarray}
& & \int d^{4}x \, \varphi_{n}^{\dagger}(x) \ast \varphi_{m}(x) 
\; (= \int d^{4}x \, \varphi_{n}^{\dagger}(x) \varphi_{m}(x))\; 
= \delta_{nm} \;. 
\end{eqnarray}
%
Under the infinitesimal transformations (3.1), 
the integration measure of the fermionic fields transforms as
\begin{eqnarray}
{\cal D}\widetilde{\psi} {\cal D}\widetilde{\bar{\psi}} = 
J_{\psi}[\alpha]{\cal D}\psi {\cal D}\bar{\psi} \;,
\end{eqnarray}
%
with the Jacobian 
\begin{eqnarray}
J_{\psi}[\alpha] \!\!&=&\!\! \det \left[\; \delta_{nm} 
 - \frac{1}{2} \alpha \int d^{4}x \, \varphi_{n}^{\dagger}(x) \ast 
\varphi_{m}(x) \; \right]^{-2} \nonumber \\
 & & \\
\!\!&=&\!\! \exp \left[\; \alpha\, \sum_{n}  
\int d^{4}x \, \varphi_{n}^{\dagger}(x) \ast 
\varphi_{n}(x)  \; \right] \nonumber \;.
\end{eqnarray}
%
In deriving the second line, we have used the identity $\ln \det 
= {\rm Tr} \ln$. In the same way as the evaluation of chiral anomalies, we 
regularize the Jacobian (3.6) with a Gaussian damping factor at hand, 
\begin{eqnarray}
J_{\psi}[\alpha] &\equiv& \lim_{\epsilon \longrightarrow 0}
\exp \left[\; \alpha \,\sum_{n} \int d^{4}x \,
\exp \left( -\;\epsilon \;\lambda_{n}^{2} \;\right) 
\varphi_{n}^{\dagger}(x) \ast \varphi_{n}(x)  \; \right] \nonumber \\
 & & \\
\!\!&=&\!\! \lim_{\epsilon \longrightarrow 0}
\exp \left[\; \alpha \,\sum_{n} \int d^{4}x \, 
\left( \;
\exp_{\ast} \left(-\; \epsilon\; /\!\!\!\!D \ast /\!\!\!\!D \right) \ast 
\varphi_{n}^{\dagger}(x) \; \right) \;
\ast \varphi_{n}(x)  \; \right] \nonumber  \;.
\end{eqnarray}
%
Here the damping factor $\exp_{\ast}$ is defined by $\exp_{\ast}x \equiv 
1+x+\frac{1}{2!}x \ast x + \cdots $. By expanding $\varphi_{n}(x)$ in plane 
waves, we can rewrite the Jacobian 
$J_{\psi}[\alpha]$ into the form 
\begin{eqnarray}
J_{\psi}[\alpha] = \exp \left[\; \alpha \,\int d^{4}x \,{\cal A}_{\psi}(x) 
\; \right] 
\nonumber \;,
\end{eqnarray}
with 
\begin{eqnarray}
\int d^{4}x \, {\cal A}_{\psi}(x) \equiv \lim_{\epsilon \longrightarrow 0}
\int d^{4}x \, \int \frac{d^{4}k}{(2\pi)^{4}} {\rm tr} \left[\;
\left( \;
\exp_{\ast}\left(-\; \epsilon\; /\!\!\!\!D \ast /\!\!\!\!D \right) \ast 
e^{ik \cdot x} \; \right) \; \ast e^{-ik \cdot x} \; \right]  \;,
\end{eqnarray}
%
where ${\rm tr}[ \quad ]$ denotes a trace over the Dirac matrices 
$\gamma^{\mu}$.  By using the identity $\gamma^{\mu}\gamma^{\nu} = 
g^{\mu\nu}+\sigma^{\mu\nu}(\equiv \frac{1}{2}[\gamma^{\mu}, 
\;\gamma^{\nu} ] )$, we obtain 
\begin{eqnarray}
& & \int d^{4}x \, {\cal A}_{\psi} \nonumber \\
& & = \lim_{\epsilon \longrightarrow 0}
\int d^{4}x \, \int \frac{d^{4}k}{(2\pi)^{4}} {\rm tr} \left[\; 
\left( \; 
\exp_{\ast}\left\{ \;-\epsilon \,\left( \; D_{\mu} \ast D^{\mu} - 
\frac{i}{2}\sigma^{\mu\nu}F_{\mu\nu}(x) \; \right) \;\right\} 
\ast e^{ik \cdot x} \;\right) \ast e^{-ik \cdot x} \; \right] \nonumber \\
 & &  \\
& & = \lim_{\epsilon \longrightarrow 0} \int d^{4}x \,
\int \frac{d^{4}k}{(2\pi)^{4}} {\rm tr} \left[\; 
\exp_{\ast}\left\{ \;-\epsilon \,\left( \; (ik_{\mu}+D_{\mu}) \ast 
(ik^{\mu}+D^{\mu}) 
- \frac{i}{2}\sigma^{\mu\nu}F_{\mu\nu}(x) \; \right) \;\right\} 
\; \right]  \;. \nonumber 
\end{eqnarray}
%
In deriving the second line of Eq. (3.9), we have utilized the fact that 
$(\partial_{\mu} e^{ik \cdot x}) \ast e^{-ik \cdot x} = ik_{\mu} 
+ \partial_{\mu}$ and $e^{ip \cdot x} \ast e^{ik \cdot x} \ast e^{-ik \cdot x} 
= e^{ip \cdot x}$. Note that the background gauge field in the covariant 
derivative $D_{\mu}$ and its field strength do not depend on the 
momentum $k_{\mu}$.

After rescaling the momentum $k_{\mu} \rightarrow k_{\mu}/\sqrt{\epsilon}$, 
we have 
\begin{eqnarray}
& & \int d^{4}x \,{\cal A}_{\psi} = 
\lim_{\epsilon \longrightarrow 0} \frac{1}{\epsilon^{2}}
\int d^{4}x \, \int \frac{d^{4}k}{(2\pi)^{4}} \,e^{k_{\mu}k^{\mu}}\,
\nonumber \\
& & \qquad \qquad 
{\rm tr} \left[ \; \exp_{\ast} \left\{ \; -2i\sqrt{\epsilon}k^{\mu} D_{\mu} 
- \epsilon D_{\mu} \ast D^{\mu} 
- \frac{i}{2}\epsilon \sigma^{\mu\nu}F_{\mu\nu}(x) \; \right\} 
\;\right]\;.
\end{eqnarray}
%
We can easily perform the momentum integration for Eq. (3.10) with the aid of 
the Gaussian integral. By using the formulas 
\begin{eqnarray}
& & \int^{\infty}_{-\infty}\frac{d^{4}k}{(2\pi)^{4}}
e^{k_{\lambda}k^{\lambda}} = \frac{1}{(4\pi)^{2}} \;, \quad 
\int^{\infty}_{-\infty}\frac{d^{4}k}{(2\pi)^{4}}
e^{k_{\lambda}k~{\lambda}} k^{\mu}k^{\nu} 
= -\frac{1}{2} \frac{1}{(4\pi)^{2}} \delta^{\mu\nu} \;, \nonumber \\
& & \\
& & \int^{\infty}_{-\infty}\frac{d^{4}k}{(2\pi)^{4}}
e^{k_{\lambda}k^{\lambda}} 
k^{\mu}k^{\nu}k^{\rho}k^{\sigma} 
= \frac{1}{4} \frac{1}{(4\pi)^{2}} 
(\delta^{\mu\rho}\delta^{\nu\sigma}
+\delta^{\mu\sigma}\delta^{\nu\rho}+\delta^{\mu\sigma}\delta^{\nu\rho}) \;,
\nonumber 
\end{eqnarray}
%
we obtain 
\begin{eqnarray}
\int d^{4}x \,
{\cal A}_{\psi}(x) = \lim_{\epsilon \longrightarrow 0} 
\frac{1}{(4\pi)^{2}}  
\int d^{4}x \,
\left( \; \frac{4}{\epsilon^{2}} 
 + \frac{2}{3} F_{\mu\nu}(x) \ast F^{\mu\nu}(x) \;\right) \;.
\end{eqnarray}
%
Here we have used the trace properties of the Dirac matrices, 
\begin{eqnarray}
{\rm tr} \mbox{\boldmath $1$} =4, \qquad 
{\rm tr}(\gamma^{\mu}\gamma^{\nu}) = 4\delta^{\mu\nu}, \qquad
{\rm tr}(\sigma^{\mu\nu}\sigma^{\rho\sigma}) =
 - 4(\delta^{\mu\nu}\delta^{\rho\sigma}+\delta^{\mu\rho}\delta^{\nu\sigma}) 
\;, \nonumber 
\end{eqnarray}
with ${\rm tr}\gamma^{\mu} = {\rm tr}(\gamma^{\mu}\gamma^{\nu}\gamma^{\rho}) 
=0$. Ignoring the term which becomes infinite in the limit 
$\epsilon \longrightarrow 0$ \cite{KF}, we obtain 
the conformal anomaly coming from the quantum effect of the fermions: 
\begin{eqnarray}
{\cal A}_{\psi}(x) =  
\frac{1}{(4\pi)^{2}}  \frac{2}{3} F_{\mu\nu}(x) \ast F^{\mu\nu}(x) \;.
\end{eqnarray}
%
%
\setcounter{subsection}{1}
\subsection{The contribution from the gauge field and ghost fields}

We next evaluate the contribution from the gauge field and the ghost fields 
to the conformal anomaly. The (global) Weyl transformation laws for the 
fluctuating field and the ghost fields are given by
\begin{eqnarray}
a_{\mu}(x) \longrightarrow \widetilde{a}_{\mu}(x) 
= \exp\left( -\alpha \right) a_{\mu}(x) \;,
\end{eqnarray}
%
%
\begin{eqnarray}
c(x)  \longrightarrow \widetilde{c}(x) 
=  \exp\left( -2\alpha \right) c(x) \;, \qquad \qquad 
\bar{c}(x) \longrightarrow \widetilde{\bar{c}}(x) 
= \bar{c}(x) \;, 
\end{eqnarray}
%
where $\alpha$ is a constant parameter. We decompose $a_{\mu}(x)$, $c(x)$, and 
$\bar{c}(x)$ as 
\begin{eqnarray}
a_{\mu}(x) = \sum_{n}c_{n}V_{\mu, n}(x) \;, 
\end{eqnarray}
%
%
\begin{eqnarray}
c(x) = \sum_{n}\alpha_{n}S_{n}(x) \;, \qquad 
\bar{c}(x) = \sum_{n}\beta_{n}S_{n}(x) \;,
\end{eqnarray}
%
respectively. Here the coefficients $c_{n}$ are the ordinary numbers and 
$\alpha_{n}$ and $\beta_{n}$ are the Grassmann numbers. For the explicit 
evaluation of the Jacobian, the basis vectors $V_{\mu, n}$ and the scalar 
basis $S_{n}$ are chosen to be the eigenfunctions, 
\begin{eqnarray}
D_{\nu}[B]D^{\nu}[B] V^{\mu}{}_{n}(x)    
-2i [\; F^{\mu\nu}[B], \; V_{\nu, n}(x) \;]_{M} 
= \lambda_{n}V^{\mu}{}_{n}(x)  \;, 
\end{eqnarray}
%
%
\begin{eqnarray}
D_{\nu}[B]D^{\nu}[B] S_{n}(x) = \lambda_{n}S_{n}(x) \;, 
\end{eqnarray}
%
with the covariant derivative defined in Eq. (2.7). We also assume that the 
sets of eigenfunctions $\{V^{\mu}{}_{n}(x)\}$ and $\{S_{n}(x)\}$ are 
orthonormal and complete, respectively.

Under the infinitesimal transformations (3.14) and (3.15), the integration 
measure of the fluctuating field and ghost fields transforms as
\begin{eqnarray}
{\cal D}\widetilde{a}_{\mu}{\cal D}\widetilde{c} {\cal D}\widetilde{\bar{c}} 
= J_{a}[\alpha] J_{c}[\alpha] {\cal D}a_{\mu} {\cal D}c {\cal D}\bar{c} \;,
\end{eqnarray}
%
with the Jacobians 
\begin{eqnarray}
J_{a}[\alpha] \!\!&=&\!\! \exp \left[\; -\alpha \sum_{n} \int d^{4}x \, 
V_{n\mu}(x) \ast V_{n}^{\mu}(x)  \; \right]  \;, \\
& & \nonumber \\
J_{c}[\alpha] \!\!&=&\!\! \exp \left[\; +2\alpha \sum_{n} \int d^{4}x \, 
S_{n}(x) \ast S_{n}(x)  \; \right]  \;.
\end{eqnarray}
%

We shall evaluate from the Jacobian (3.22) first. For convenience, let us 
introduce the notation with respect to the covariant derivative 
$D_{\mu}S_{n}(x) 
\equiv \int\frac{d^{4}k}{(2\pi)^{4}}D_{\mu}[;k]\hat{S}_{n}(k)
e^{ik \cdot x}$, where 
\begin{eqnarray}
D_{\mu}[;k]
\equiv  \partial_{\mu}(=ik_{\mu}) - 2\int\frac{d^{4}p}{(2\pi)^{4}} 
\hat{B}_{\mu}(p)e^{ip \cdot x} 
\sin\left( \frac{1}{2}p \wedge k \right)  \;,
\end{eqnarray}
%
with $p \wedge k \equiv p_{\rho}\theta^{\rho\sigma}k_{\sigma}$. 
The background gauge field in the covariant derivative $D_{\mu}$ 
depends on the momentum $k_{\mu}$ via the sine functions 
$\sin\left( \frac{1}{2}p \wedge k \right)$, since the covariant 
derivative contains the Moyal bracket. As we shall see, the sine function 
$\sin\left( \frac{1}{2}p \wedge k \right)$ corresponds to 
the structure constants in ordinary gauge theories. 
The covariant derivatives 
$D_{\mu}[;k]$ satisfy the following commutation relation: 
\begin{eqnarray}
[\, D_{\mu}[;k], \; D_{\nu}[;k] \,] = -iF_{\mu\nu}(x; k) \equiv 
\int \frac{d^{4}p}{(2\pi)^{4}} \hat{F}_{\mu\nu}(p)e^{ip \cdot x} 
(-2)\sin \left( \frac{1}{2}p \wedge k \right) \;, 
\end{eqnarray}
%
where $\hat{F}_{\mu\nu}(p)$ is the Fourier transformation of $F_{\mu\nu}$: 
$F_{\mu\nu}(x) \equiv \int \frac{d^{4}p}{(2\pi)^{4}} 
\hat{F}_{\mu\nu}(p)e^{ip \cdot x}$. In deriving this expression, we have 
used the relations \\
$\sin\left(p \wedge (q+k) \right)  
\sin\left(q \wedge k \right)  
 - \sin\left(q \wedge (p+k) \right)  
 \sin\left(p \wedge k \right)  
= \sin\left(p \wedge q \right)  
\sin\left( (p+q) \wedge k \right)$.

Under the Fourier transformations for $S_{n}(x)$, the Jacobian (3.22) takes 
the following form: 
\begin{eqnarray}
J_{c}[\alpha] = \exp \left[\; 
+2\alpha \sum_{n} \int d^{4}x \, 
\int\frac{d^{4}k}{(2\pi)^{4}} \int\frac{d^{4}l}{(2\pi)^{4}}
\hat{S}_{n}(k)  \hat{S}_{n}(l)  \; \right]  \;.
\end{eqnarray}
%
In the same way as Eq. (3.7), we regularize the Jacobian 
(3.25) with a Gaussian damping factor as 
\begin{eqnarray}
J_{c}[\alpha] = \exp \left[\; 
\alpha \int d^{4}x \, {\cal A}_{c}(x) \; \right] \;, \nonumber 
\end{eqnarray}
with 
\begin{eqnarray}
& & \int d^{4}x \,{\cal A}_{c}(x) \equiv 2\lim_{\epsilon \longrightarrow 0}
\sum_{n} \int d^{4}x \, 
\int\frac{d^{4}k}{(2\pi)^{4}} \int\frac{d^{4}l}{(2\pi)^{4}}
\left(\; \exp\left(-\; \epsilon\; D_{\mu}[;k] D^{\mu}[;k] \;\right)
\hat{S}_{n}(k) \; \right) \hat{S}_{n}(l)  \nonumber \\ 
& & = 2\lim_{\epsilon \longrightarrow 0} \int d^{4}x \, 
\int\frac{d^{4}k}{(2\pi)^{4}} 
\exp\left\{-\; \epsilon\; (ik_{\mu}+D_{\mu}[;k]) 
(ik^{\mu}+D^{\mu}[;k]) \right\} \;.
\end{eqnarray}
%
In deriving Eq. (3.26), we have used the identity $\partial_{\mu}
\hat{S}_{n}(k) = ik_{\mu}\hat{S}_{n}(k)$. Note that the 
background gauge field depends on the momentum $k_{\mu}$. This means that the 
Jacobian factor (3.26) includes the {\it nonplanar} contribution 
\cite{VVKGT, SMMVR}.
In terms higher than the second power of the covariant derivative, 
however, we can decompose them into terms depending on the momentum 
$k_{\mu}$, and terms independent of the momentum $k_{\mu}$. 
In the fourth power of the covariant derivative, for example, we obtain  
\begin{eqnarray}
& & \int d^{4}x \, \biggl[\; 
D_{[\mu}[;k] \; D_{\nu]}[;k] \; 
D_{[\rho}[;k] \; D_{\sigma]}[;k] \; \biggr] \\
& & = 4 \int d^{4}x \, \biggl( 
\int \frac{d^{4}p}{(2\pi)^{4}} \hat{F}_{\mu\nu}(p)e^{ip \cdot x} 
\sin \left( \frac{1}{2}p \wedge (q+k) \right) \, 
\int \frac{d^{4}q}{(2\pi)^{4}} \hat{F}_{\rho\sigma}(q)e^{iq \cdot x} 
\sin \left( \frac{1}{2}q \wedge k \right) 
\biggr)  \nonumber \\
& & = -2\int d^{4}x \, \biggl( 
\int \frac{d^{4}p}{(2\pi)^{4}} \hat{F}_{\mu\nu}(p)e^{ip \cdot x} 
\int \frac{d^{4}q}{(2\pi)^{4}} \hat{F}_{\rho\sigma}(q)e^{iq \cdot x} 
\biggr) \; + (\mbox{\rm the term that depends on $k_{\mu}$}) 
\nonumber \;,
\end{eqnarray}
%
under the integration with respect to the space-time coordinates 
$x^{\mu}$. Here $D_{[\mu}D_{\nu]} \equiv D_{\mu}D_{\nu}-D_{\nu}D_{\mu}$. 
In deriving the third line of Eq. (3.27), we have made use of the Fourier 
inverse transform for the delta function $\displaystyle{ \int d^{4}x \, 
e^{i(p+q) \cdot x} = (2\pi)^{4}\delta(p+q) }$ and an identity 
\begin{eqnarray}
\sin^{2}(\frac{1}{2} p \wedge k) = 
\frac{1}{2}(1-\cos(p \wedge k)) \;.
\end{eqnarray}
%
We find that the integral over space-time 
coordinates plays a role corresponding to a trace about the generators of 
gauge groups in ordinary non-Abelian gauge theories. In order for the 
integration to play the role of the trace, however, note that the 
infinitesimal parameter 
$\alpha$ in the Weyl transformation must be a constant. The first term in the 
third line of Eq. (3.27) can be regarded as the {\it planar} contribution 
\cite{VVKGT, AMLSNT, MMSJ}. Such a {\it planar} contribution can be 
expressed as 
\begin{eqnarray}
\int d^{4}x \left. \biggl(\; 
D_{[\mu}[;k] \; D_{\nu]}[;k] \; 
D_{[\rho}[;k] \; D_{\sigma]}[;k] \; \biggr) 
 \right|_{planar} 
= -2\int d^{4}x \, F_{\mu\nu}(x) \ast F_{\rho\sigma}(x) \;.
\end{eqnarray}
%
%
Here we have used the fact that $\displaystyle{\int d^{4}x \, f(x) \ast g(x) 
= \int d^{4}x  
\int \frac{d^{4}p}{(2\pi)^{4}} \hat{f}(p)e^{ip \cdot x} 
\int \frac{d^{4}q}{(2\pi)^{4}} \hat{g}(q)e^{iq \cdot x}}$. 
Contributions from the second and the third power of the covariant derivative 
in the planar sector cancel after the momentum integration. It is the same 
in the fourth power of the covariant derivative with the symmetric property 
for the Minkowski indices.  
Therefore, the momentum integration in the planar sector leads to the 
following result: 
\begin{eqnarray}
& & \int d^{4}x \, \left. {\cal A}_{c}(x) \right|_{planar} \nonumber \\
& & = 2\lim_{\epsilon \longrightarrow 0} \frac{1}{\epsilon^{2}}
\int d^{4}x \, \int \frac{d^{4}k}{(2\pi)^{4}} \,e^{k_{\mu}k^{\mu}} 
\left. \exp \left\{ \; -2i\sqrt{\epsilon}k^{\mu} D_{\mu}[;k] 
- \epsilon D_{\mu}[;k]D^{\mu}[;k] \; \right\} \right|_{planar} 
\nonumber \\ 
& & = \lim_{\epsilon \longrightarrow 0} \frac{1}{(4\pi)^{2}} 
\int d^{4}x \, \left( \; \frac{1}{\epsilon^{2}} 
 - \frac{2}{6}F_{\mu\nu}(x) \ast F^{\mu\nu}(x) \;\right) \;. 
\end{eqnarray}
%
Here we have made use of the formulas (3.11). Ignoring the infinite term, we 
obtain the conformal anomaly from the quantum effect of the ghost fields, 
\begin{eqnarray}
\left. {\cal A}_{c}(x) \right|_{planar} =  
\frac{1}{(4\pi)^{2}}  \left( -\frac{1}{3} \right) 
F_{\mu\nu}(x) \ast F^{\mu\nu}(x) \;.
\end{eqnarray}
%

We next evaluate the Jacobian (3.21). Under the Fourier transformations for 
$V_{n}^{\mu}(x)$, the Jacobian (3.21) takes the following form: 
\begin{eqnarray}
J_{a}[\alpha] = \exp \left[\; 
+2\alpha \sum_{n} \int d^{4}x \, 
\int\frac{d^{4}k}{(2\pi)^{4}} \int\frac{d^{4}l}{(2\pi)^{4}}
\hat{V}_{n\mu}(k)  \hat{V}_{n}^{\mu}(l)  \; \right]  \;,
\end{eqnarray}
%
where we have used the relations 
$\sum_{n}\hat{V}_{n\mu}(k)\hat{V}_{n}^{\mu}(l) = (2\pi)^{4}
\delta^{4}(k+l)$. Taking account of Eq. (3.18), we regularize the Jacobian 
(3.32) as follows: 
\begin{eqnarray}
& & J_{a}[\alpha] = \exp \left[\; \alpha \int d^{4}x \, {\cal A}_{a}(x) 
\; \right] \;, \nonumber 
\end{eqnarray}
with 
\begin{eqnarray}
& & \int d^{4}x \, {\cal A}_{a}(x) \nonumber \\
& & \equiv -\lim_{\epsilon \longrightarrow 0} 
\sum_{n} \int\frac{d^{4}k}{(2\pi)^{4}} \int\frac{d^{4}l}{(2\pi)^{4}}
\left( \;\exp\left\{-\; \epsilon\; \left(\; 
\delta_{\mu}{}^{\nu}D^{2}[;k] -2iF_{\mu}{}^{\nu}(x ;k) \;\right) \;\right\}
\hat{V}_{n \nu}(k) \; \right) \hat{V}_{n}^{\mu}(l) \nonumber \\ 
& & = -\lim_{\epsilon \longrightarrow 0} \int d^{4}x \, 
\int\frac{d^{4}k}{(2\pi)^{4}} 
{\rm tr} \left[\; \exp\left\{-\; \epsilon\; \left(\; 
\delta_{\mu}{}^{\nu}(ik + D[;k])^{2} 
-2iF_{\mu}{}^{\nu}(x; k) \;\right) \;\right\} \; \right] \;,
\end{eqnarray}
%
where $D^{2} \equiv D_{\mu}D^{\mu}$ and ${\rm tr}[ \quad ]$ denotes a trace 
with respect to the Minkowski indices. The concrete form of the field strength 
$F_{\mu\nu}(x; k)$ is shown in Eq. (3.24). Since the background gauge field 
and its field strength depend on the momentum $k_{\mu}$, the Jacobian factor 
(3.33) also includes the {\it nonplanar} contribution. Selecting the 
{\it planar} contribution, we arrive at 
\begin{eqnarray}
& & \int d^{4}x \, \left. {\cal A}_{a}(x) \right|_{planar} \nonumber \\
& & = -\lim_{\epsilon \longrightarrow 0} \frac{1}{\epsilon^{2}}
\int d^{4}x \, \int \frac{d^{4}k}{(2\pi)^{4}} \,e^{k_{\mu}k^{\mu}} 
\left. {\rm tr} \left[\; \exp \left\{ \; 
-2i\sqrt{\epsilon}\delta_{\mu}{}^{\nu} k^{\lambda} D_{\lambda}[;k] 
- \epsilon \delta_{\mu}{}^{\nu}D^{2}[;k] 
+ 2i F_{\mu}{}^{\nu}(x; k) \; \right\} \right] \right|_{planar} 
\nonumber \\ 
& & = \lim_{\epsilon \longrightarrow 0} \frac{1}{(4\pi)^{2}} 
\int d^{4}x \, \left( \; -\frac{1}{\epsilon^{2}} 
 - \frac{5 \times 2}{3}F_{\mu\nu}(x) \ast F^{\mu\nu}(x) \;\right) 
\end{eqnarray}
%
Here we have made use of the formulas (3.11). Ignoring the infinite term, we 
obtain the conformal anomaly from the quantum effect of the fluctuating 
field, 
\begin{eqnarray}
\left. {\cal A}_{a}(x) \right|_{planar} =  
\frac{1}{(4\pi)^{2}}  \left( -\frac{10}{3} \right) 
F_{\mu\nu}(x) \ast F^{\mu\nu}(x) \;.
\end{eqnarray}
%

%
\setcounter{subsection}{2}
\subsection{The conformal anomaly and the $\beta$ function}

Adding together the contributions from matter fields, the gauge field, and 
ghost fields, we obtain the conformal anomaly in noncommutative QED. The 
explicit form of the conformal anomaly is given by 
\begin{eqnarray}
& & \int d^{4}x \, \left. {\cal A}(x) \right|_{planar} 
= \int d^{4}x \, \left( \,  n_{f} \;\cdot\;{\cal A}_{\psi}(x) 
+ \left. {\cal A}_{a}(x) \right|_{planar} 
+ \left. {\cal A}_{c}(x) \right|_{planar} \, \right) \nonumber \\
& & = \frac{1}{(4\pi)^{2}}  \left( \frac{2}{3}n_{f} - \frac{11}{3} \right) 
\int d^{4}x \, F_{\mu\nu}(x) \ast F^{\mu\nu}(x) \;,
\end{eqnarray}
%
where $n_{f}$ is the number of flavors. This takes the same form as 
the conformal anomaly in ordinary QED except for the ordinary product replacing the Moyal star product. 
The gauge invariance of the result is guaranteed by 
the integration over space-time coordinates. Note that the coefficient of 
the conformal anomaly differs from the coefficient of the conformal 
anomaly in ordinary QED by a factor of 2 \cite{KF}. The difference 
comes from the identity (3.28) and normalization of the $U(1)$ generator. 

A close relation exists between the conformal anomaly and the $\beta$ function 
in the ordinary field theories. When quantum corrections are 
included, a scale transformation shifts the renormalized coupling constant. 
Since the variation is proportional to the $\beta$ function, the corresponding 
change in the action is also proportional to the $\beta$ function in a 
classically scale invariant theory. Therefore, the conformal anomaly is 
proportional to the $\beta$ function. In ordinary QED with massless fermions, 
the conformal anomaly up to the one-loop correction is given by 
\begin{eqnarray}
\int d^{4}x \,{\cal A}(x) 
= \frac{\beta(e)}{2e^{3}} \int d^{4}x \, 
F_{\mu\nu}(x)F^{\mu\nu}(x) \;,
\end{eqnarray}
%
where $e$ denotes the coupling constant. This expression shows the relation 
between the conformal anomaly and $\beta$ functions in ordinary QED. 
In analogy with the ordinary QED, we express the conformal anomaly 
in noncommutative QED as follows: 
\begin{eqnarray}
\int d^{4}x \,{\cal A}(x) 
= \frac{\left. \beta(g) \right|_{NC-QED}}{2g^{3}} 
\int d^{4}x \, F_{\mu\nu}(x) \ast F^{\mu\nu}(x) \;,
\end{eqnarray}
%
where $g$ is the coupling constant in noncommutative QED. From Eqs. (3.36) 
and (3.38), we can evaluate the $\beta$ function in noncommutative QED up to 
the one-loop contribution:
\begin{eqnarray}
\left. \beta(g) \right|_{NC-QED} = -\frac{g^{3}}{(4\pi)^{2}}
\left(\;\frac{22}{3} - \frac{4}{3}n_{f} \;\right) \;. 
\end{eqnarray}
%
This is coincident with the $\beta$ function obtained from the perturbative 
analysis \footnotemark[2] \cite{MH}.  
\footnotetext[2]{By the definition of a $\beta$ function, the 
expression (3.39) is different from that in Ref. \cite{MH} only in 
the factor $\frac{1}{g}$.  
}
%

%
%
\section{The generalization to the $U(N)$ gauge group}
\setcounter{equation}{0}
\addtocounter{enumi}{1}

It is straightforward to modify the method of the calculation in the $U(1)$ 
gauge group performed for the previous section to the $U(N)$ gauge group. 
In the $U(N)$ gauge group, the notation ${\rm tr}[ \quad ]$ in Eq. (3.8) 
replaces the notation denoting the trace over the Dirac matrices 
$\gamma^{\mu}$ and the gauge group generators $T_{a}$ with a certain 
irreducible representation. Similarly, the trace over the gauge group 
generators with the adjoint representation appears in Eq. (3.26) and the 
notation ${\rm tr}[ \quad ]$ in Eq. (3.33) replaces the notation 
denoting the trace with respect to the Minkowski indices and the trace over 
the gauge group generators with the adjoint representation. Hence, the 
factor $C(r)$ caused from the normalization ${\rm tr}[T_{a}T_{b}]
=C(r)\delta_{ab}$ appears in the coefficient of the corresponding 
expression for Eq. (3.13), and the factor $C_{2}(G) (= N )$ from the quadratic 
Casimir operator for the $U(N)$ gauge group appears in the coefficient of the 
corresponding expression for Eqs. (3.31) and (3.35). The corresponding 
conformal anomaly in noncommutative QCD with the $U(N)$ gauge group is given 
as follows: 
\begin{eqnarray}
\int d^{4}x \, \left. {\cal A}(x) \right|_{planar} 
= \frac{1}{(4\pi)^{2}}  \left( \frac{2}{3}n_{f}C(r) 
- \frac{11}{3}C_{2}(G) \right) 
\int d^{4}x \, F_{\mu\nu}{}^{a}(x) \ast F^{\mu\nu}{}_{a}(x) \;,
\end{eqnarray}
%
where $F_{\mu\nu}{}^{a}$ are the components of the field strength:
$F_{\mu\nu} = F_{\mu\nu}{}^{a}T_{a}$. Note also that the coefficient of 
the conformal anomaly differs from the coefficient of the conformal 
anomaly in ordinary QCD with the $U(N)$ gauge group. When evaluating the 
$\beta$ function based on the relation corresponding to Eq. (3.38) in 
noncommutative QCD, we obtain 
\begin{eqnarray}
\left. \beta(g) \right|_{NC-QCD} = -\frac{g^{3}}{(4\pi)^{2}} \; 
2\left(\;\frac{11}{3}C_{2}(G) - \frac{2}{3}n_{f}C(r) \;\right) \;.
\end{eqnarray}
%
This is also coincident with the $\beta$ function obtained from the 
perturbative analysis \footnotemark[3] \cite{VVKGT}. 
\footnotetext[3]{If we take a different normalization of a $U(1)$ generator, 
we might also obtain the one-loop $\beta$ function in ordinary 
$U(N)$ gauge theory. For example, see Ref. \cite{AA}.}
%

%
%
\section{Conclusion and Discussion}
\setcounter{equation}{0}
\addtocounter{enumi}{1}

In this paper, we have evaluated the conformal anomalies in noncommutative 
gauge theories. As well as the axial anomalies and chiral gauge anomalies, 
the conformal anomalies are also calculable on the basis of the path integral 
formulation. Except for the difference in coefficients, the conformal 
anomalies (3.36) and (4.1) take the form of the straightforward Moyal 
deformation in the corresponding anomalies in ordinary gauge theories. 
The difference in coefficients is a consequence of the noncommutativity of the 
Moyal star product. 
In evaluating the conformal anomalies by the path integral formulation, 
the background field method has been adopted. The fluctuating fields and 
the ghost fields in the background field method transform as a field in the 
adjoint representation under the gauge transformation (2.13). 
The interactions with the background field and such adjoint fields include 
the Moyal bracket. Then the sine function $\sin(\frac{1}{2}p \wedge k)$ 
arising from the Moyal bracket plays the role of the structure constants in 
the gauge algebra. The corresponding quadratic Casimir operator takes the 
value of 2. It causes the difference in coefficients of the conformal 
anomalies between the ordinary gauge theory and noncommutative gauge theory. 

There is a relation between the conformal anomaly and the $\beta$ function 
in ordinary gauge theories. The $\beta$ function in noncommutative gauge 
theory can be evaluated by applying the relation between the conformal anomaly 
and the $\beta$ function. The evaluation of the $\beta$ function based on this 
relation is in agreement with the result of the perturbative analysis. 

In evaluating the conformal anomalies, we have focused attention on the 
planar contribution. We can confirm that the nonplanar sector in the first 
and second power in Taylor series of the Gaussian damping factor does not 
contribute in the case of 
$p \circ p \equiv p_{\mu}\theta^{\mu\rho}p^{\nu}\theta_{\nu\rho} < 0$. 
We shall report elsewhere whether the nonplanar sector does not 
contribute to the conformal anomalies in all order in Taylor series of the 
Gaussian damping factor.

We have confined our discussion to the conformal anomaly under the global 
Weyl transformation. If we extend the global Weyl transformation to the local 
Weyl transformation, then the integration over space-time coordinates does not 
play the role corresponding to the trace over the generators of gauge groups 
in ordinary non-Abelian gauge theories. 
Further consideration is needed in this respect. 
In the noncommutative field theory at the classical level, the 
variation of the action under the global scale (or dilatation) transformation 
for the fields can be expressed as the variation of the action under the 
global scale transformation of the noncommutativity 
parameter \cite{AGJGHGLPMSRW}. 
In order to argue about the conformal anomaly under the local Weyl 
transformation, it may be necessary to consider the noncommutative field 
theory with nonconstant noncommutative parameters \cite{PMHSPM}. 
We hope to discuss this subject in the future.


\section*{Acknowledgments}

I am grateful to S. Deguchi for valuable discussions and helpful suggestions. 
I would like to thank A. Sugamoto for a careful reading of the manuscript 
and useful comments. I also would like to thank A. Armoni, J. Gomis, and 
R. Schiappa for useful comments.

\end{document}